**Özet**

*Bu çalışma, endüstriyel iş sahalarında farklı risk kategorisine sahip çalışma bölgelerinde istihdam edilen bakım personelinin güvenliklerini sağlamak amacıyla, ilgili çalışan ve çevre şartları hakkında bilgi toplayan bir giyilebilir cihazın geliştirilmesini amaçlamaktadır. Bu cihaz, iş kazalarını önlemek için bakım personelinden toplanan sağlık parametrelerine göre kişinin stres seviyesini ölçerek ilgili risk seviyesine ait bakım için uygun olup olmadığı anlık olarak değerlendirebilir ve gerektiği durumlarda üzerinde çalıştığı makineyi acil pozisyonuna getirebilir. Bu çalışma kapsamında geliştirilen sistem dört bölümden oluşmaktadır. Giyilebilir ünite, akıllı telefon, server bilgisayar ve makine üzerine monte edilen kontrol ünitesi. Giyilebilir cihaz, takılı olduğu bakım personelinin nabız, oksijen saturasyonu, vücut sıcaklığı ve deri direncini ölçebilmekte ayrıca ilgili personelin çalıştığı ortama ait; ısı, ışık, nem, CO değerlerinin ölçümü yapılabilmektedir. Bu verilerin bir karar destek sistemi algoritmasına girişi yapılarak elde edilen sonuç ile makinenin acil durumuna geçirilmesi ya da ilgili operatörün bu bakım için uygunluğunun öngörülmesi sağlanacaktır. Bu uygulama önemli bir kişisel güvenlik sistemidir ve bakım ve/ya onarım işlemleri sırasında meydana gelebilecek iş kazalarını önleyebilir.*

**Anahtar kelimeler:** Giyilebilir Elektronik, Güvenlik, İş Sağlığı ve Güvenliği, Güvenlik Cihazı, Nesnelerin İnterneti

**Abstract**

*This study aims to develop a wearable device that collect health data from maintenance personnel and environmental conditions' data in order to ensure the safety of the staff in industrial work areas where have different levels of risk categories. This device can stop machine immediately according to the health parameters that collected from the maintenance personnel to prevent work accidents and also with measuring stress level of this personnel, system make prediction of suitability of personnel for maintenance operation according to degree of risk level of machine. System consists from four parts. Wearable unit, smartphone, server computer and machine mounted unit. The wearable device can measure personnel; heart rate, oxygen saturation, body temperature, skin resistance; and also collect data from environment; heat, light, humidity, CO2.. The results which obtained by decision support system algorithm can stop machine immediately or predict the suitability of corresponding operator for this maintenance operation. This application is an important personal safety system and can prevent work accidents that may occur during maintenance or repair operations.*

# İş kazalarını önleyebilmek için bakım/onarım personelinin kullanabileceği bir giyilebilir emniyet sisteminin tasarlanması

# Design of a Wearable Safety System for Maintenance Staff to Prevent Work Accidents

Ersin BERBEROĞLU[1]
Mahmut TOKMAKÇI[2]
Ahmet Turan ÖZDEMİR[3]

[1]Arge Merkezi, Merkez Çelik A.Ş, Kayseri, Türkiye,
[2]Biyomedikal Mühendisliği Bölümü, Erciyes Üniversitesi, Kayseri, Türkiye
[3]Elektrik Elektronik Mühendisliği Bölümü, Erciyes Üniversitesi, Kayseri, Türkiye

## 1. Giriş

Günümüzde endüstriyel tesisler, makineleşmenin artması ve çalışma alanlarını büyümesiyle gittikçe daha karmaşık bir hal almaktadır. Bu duruma paralel olarak, çalışan üzerinde oluşan üretim baskısı ve yeni teknolojilere adaptasyon sürecinde, istenmeyen iş kazaları meydana gelebilmektedir. TUİK'in (Türkiye İstatistik Kurumu) verilerine göre Türkiye'deki iş kazalarının %51'den fazlası üretim sektöründe yaşanmıştır. 2011 senesinde 692227 adet iş kazasında 1700 kişi olan ölüm sayısı 2016 senesinde 286068 adet iş kazasında 1405 kişi olarak kayıtlara geçmiştir [1].

**Tablo 1.** Türkiye'de Meydana Gelen İş Kazaları ve Ölümleri

| | 2011 | 2012 | 2013 | 2014 | 2015 | 2016 |
|---|---|---|---|---|---|---|
| Toplam iş kazası | 69.227 | 74.871 | 191.389 | 221.366 | 241.547 | 286.068 |
| Meydana gelen ölüm | 1700 | 744 | 1360 | 1626 | 1252 | 1405 |

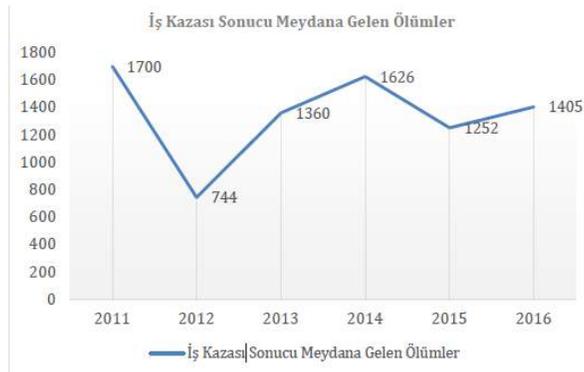

**Şekil 1.** Türkiye'de İş Kazası Sonucu Meydana Gelen Ölümlerin Seyri

Ölümcül kazaların yaşandığı günümüzde, iş sağlığı ve güvenliği çalışanlar ve işverenler için oldukça önem kazanmıştır. İş kazalarını önlemek için sebeplerini anlamak oldukça önemlidir. İş kazalarının başlıca sebepleri;

• Kişisel güvenlik ekipmanlarının kullanılmaması veya olmaması,

• Çalışma esnasında uykulu, gergin, hasta veya dalgın olma

• Çalışma ortamının tertipsiz ve düzensiz olması,

• Çalışma ortamında iş haricinde gereksiz hareketler yapmak.

Şekil 2'de TS EN 13849 harmonize standardında verilen endüstriyel ortamlarda meydana gelebilecek iş kazalarının risk grafiği verilmiştir [2]. Buna göre en yüksek risk S2, F2 ve P2 şartlarında oluşmaktadır. Grafikte Y harfi ile gösterilmiştir. En düşük risk ise S1,F1,P1 durumlarında oluşmaktadır.

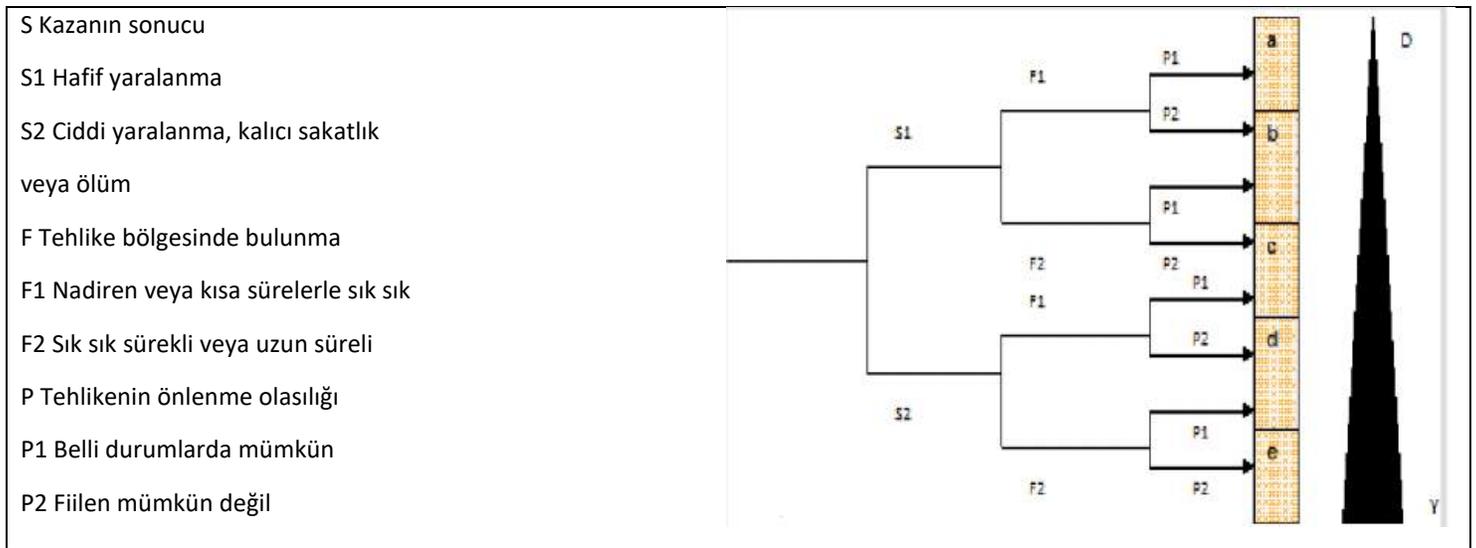

| S Kazanın sonucu |
| --- |
| S1 Hafif yaralanma |
| S2 Ciddi yaralanma, kalıcı sakatlık veya ölüm |
| F Tehlike bölgesinde bulunma |
| F1 Nadiren veya kısa sürelerle sık sık |
| F2 Sık sık sürekli veya uzun süreli |
| P Tehlikenin önlenme olasılığı |
| P1 Belli durumlarda mümkün |
| P2 Fiilen mümkün değil |

**Şekil 2.** İş kazaları risk grafiği

Bu kapsamda personelin kullanabileceği bir giyilebilir emniyet kiti tasarlayarak, tehlikeli iş kazalarının sayısının düşürülmesi amaçlanmıştır. Giyilebilir ürünlerin günümüzde akıllı bileklik, akıllı saat, akıllı gözlük, akıllı giysiler gibi pek çok ürün şeklinde yaygınlaştığı görülmektedir. Akıllı giyilebilir sistemlerin (SWS, Smart Wearable System) çeşitleri Chan ve arkadaşlarının yapmış oldukları çalışmada detaylı olarak sınıflandırılmıştır [3]. Kişilerin sağlık parametrelerine ulaşabilmek için hem giyilebilir hem de implant edilebilir teknolojiler kullanılabilmektedir. Bu cihazlar içerdikleri farklı yapılardaki sensörler, akıllı tekstiller, güç kaynakları, kablosuz haberleşme donanımları, işlemciler, multimedya cihazları, kullanıcı arayüzleri, yazılımlar ve karar destek sistemleri kullanarak vücut ve deri sıcaklığı, kalp atışı, kalp atış değişkenliği, kan basıncı, kandaki oksijen seviyesi, EKG, EEG ve solunum hızı gibi hayati sinyallerin izlenebilmesi ve yorumlanmasına olanak sağlamaktadır.

Giyilebilir akıllı sistemler kişisel sağlık parametrelerini elde etmede en etkin araç olarak gösterilmektedir. Bireylerden biyomedikal veri toplamak amacı ile akıllı, düşük enerji tüketimli, hafif ve ucuz sensör ve ağ yapıları üzerine çalışmalar günümüzde yoğun bir şekilde yürütülmektedir [4-7].

Giyilebilir akıllı sistemlerin kişisel güvenlik amacı ile kullanımı ağırlıklı olarak merkez dışında çalışan servis personelinin ve maden işçilerinin kullanımına yönelik geliştirildiği görülmektedir. Bu cihazların çalışan personelin hareket kabiliyeti ve el kullanma yeteneğini kısıtlamayacak şekilde bilek ve kafa bölgesine takılabilecek yapılarda ya da giyilebilir bir kıyafet şeklinde tasarlandıkları görülmektedir. Aleksy ve arkadaşları ABB firması öncülüğünde bir Giyilebilir Emniyet Kıyafeti geliştirmişlerdir [8]. Geliştirdikleri kıyafet üzerinde ortam sıcaklığını ölçen bir sensör, vücut sıcaklığını ölçen bir sensör, ortamdaki karbonmonoksiti ölçen sensör, nem sensörü, nabız tespit sensörü ve acil yardım butonu bulunmaktadır. Servis personelinde acil bir durum oluştuğunda giyilebilir kıyafet üzerindeki kontrolcü Bluetooth haberleşme arayüzü üzerinden mobil bir cihaza ihbar göndermekte ve bu bildirim GPS konum bilgisi ile birleştirilerek izleme merkezine iletmektedir. Kıyafet içerisinde sensörlerlere erişim için vücut alan-ağı (BAN Body Area Network) protokolü kullanılmıştır. Bu çalışmada kontrolcü olarak Ardunio Lily Pad kullanılmış ve C yazılım dili kullanılarak programlanmıştır. Acil durum anında merkez istasyona mobil telefon üzerinden şebeke ağı aracılığı ile kısa mesaj bildirimi ya da doğrudan arama yapılmaktadır.

Giyilebilir Akıllı Sistemlerin kullanılabilmesi için yoğun araştırma yapılan sektörlerden biri de inşaat sektörüdür. Endüstriyel sektörler içinde inşaat ölümcül iş kazası, sakat kalma ve hastalık kapma riskinin en yüksek olduğu meslek grubudur. İnşaat sektörü için kurgulanacak güvenlik sistemlerinde sadece fiziksel sinyallerin alınması yeterli değildir. Aynı zamanda "hands-free" eller serbest sistemler ile çalışma alanından sürekli veriler alınarak çevresel koşulların analiz edilmesi ve çalışanın tehlikeli bölgelere yaklaştığında alarm vermesinin sağlanması da gerekmektedir. Avolusi ve arkadaşlarının yapmış oldukları tabloda inşaat alanında olabilecek kazaları gruplamışlar ve bu kazalara karşı hangi parametrelerin takip edilmesi gerektiğini vurgulamışlardır [9]. Yüksekten düşme, kayma ve takılma gibi oluşabilecek iş kazalarının tespitinde fizyolojik sinyallerin izlenmesinin gerektiği, belirtilmiştir. Bu sinyaller kalp atışı, nefes hızı, vücut pozisyonu, vücut hızı, vücut yönü ve vücut sıcaklığı, yürüme adım sayısı olarak verilmiştir. Çevresel şartların etmen olabileceği iş kazalarında (patlama ve yangın gibi) ortam sıcaklığı, nemi, basıncı, sesi ve hava kalitesinin izlenmesi gerektiği, hareketli objeler arasında sıkışma gibi kazaların önüne geçmede ise çalışanın anlık pozisyonunu izleme, materyalleri ve hareketli araçları takip etme gibi özelliklerin olması gerektiğini vurgulamışlardır [10].

Lee ve Chung çalışmalarında araç sürücülerinin dikkatlerinin dağıldığında ve uykusu geldiğinde uykunun şiddetine göre beş kademeli olarak uyarı veren bir sistem üzerine çalışmışlardır. Arduino Lily pad ile Fotopletismografik ve Galvanik Deri Direnç sensörü kullanarak kişinin stres seviyesini ölçme temelli çalışan bir yapı kurgulamışlardır. Sisteme Bluetooth düşük enerji (BLE Bluetooth Low Energy) birimi entegre ederek mevcut akıllı telefonlar ile haberleşmesini mümkün kılmış ve kişinin nabız hızı, nabız hızı değişkenliği, stres seviyesi ve nefes hızı bilgileri transfer etmişleridir. Çalışmadaki karar destek yazılımı mobil aplikasyon üzerinden koşturulmuş, uyarılar akıllı telefon üzerinden verilmiştir [11]. Zilberg ve arkadaşları sürücü koltuklarına yerleştirdikleri sistemler ile kişinin uyanıklık durumunu tespit etmeye çalışmışlarıdır [12]. Araç uyarıcı sistemler arasında direksiyon üzerine montaj edilen yapılar da literatürde yer almaktadır [13]. Sano ve Pikard bu çalışmalara ek olarak sürüş işleminden önce kişiye anket girişi yaptırarak uyku ve kahve alımı gibi parametreleri de girdi olarak irdelemişlerdir [14].

## 2. Yöntem ve Metotlar

Giyilebilir Güvenlik Kiti Sistemi Projesi başlangıçta iki birimden tasarlanmış ancak ilerleyen süreçlerde artan ihtiyaçlar ve eklenmek istenen özellikler sebebiyle sistem bileşenleri artmıştır. Başlangıçta noktadan-noktaya (M2M) haberleşmesi temelli düşünülen sistemde bileşenler bir alıcı ve vericiden meydana gelmekteydi. Bu haliyle alıcı devre, makine kontrolcü birimi ile eşzamanlı çalışacak olan kısımdan müteşekkil iken verici kısım, nabız, oksijen saturasyonu ve jiroskop verilerini sağlayan giyilebilir cihazdan oluşmaktaydı [15].

Projenin gelinen aşamasında sistem dört birimden oluşmaktadır. Giyilebilir cihaz birimi görevi itibariyle başlangıçta düşünülen şekli ile aynı kapsamdadır ancak kullandığı sensörler farklılaştırılmıştır. Operatörün stres, heyecan gibi duygusal durumlarının tespiti amacı ile deri direnç sensörü ve ortamın sıcaklık, nem, ışık ve $CO_2$ değerlerini ölçebilecek bir multi sensör sisteme dahil edilmiştir. Sensörlerden toplanan bu veriler hem giyilebilir cihazda kullanılan OLED ekrana aktarılmakta hem de Bluetooth haberleşmesi ile Android tabanlı mobil cihaza gönderilmektedir. Android cihazda bulunan dâhili jiroskop ve ivmeölçer diğer sensör verileriyle birleştirilerek kablosuz internet ağı üzerinden MQTT (Message Queuing Telemetry Transport) protokolü kullanılarak tüm sistemin kontrolünün ve takibinin yapılacağı bir ana bilgisayara iletilmektedir.

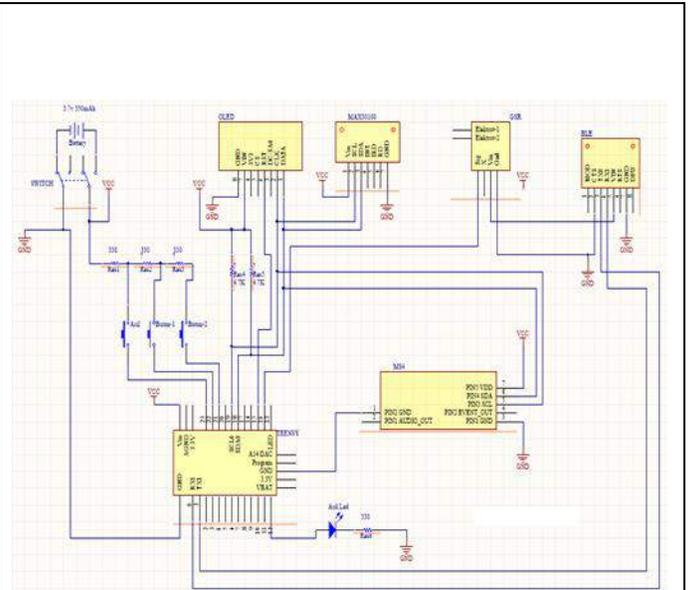

**Şekil 3.** Giyilebilir Cihaz Örnek Şematik

Operatöre ve ortama ait ilgili verilerin toplanmasında kullanılan sensörler sırasıyla, MAX30100 Galvanik Deri Tepki sensörü ve MS4 Multi-Sensör modülüdür. Sistemin bileşenleri ise OLED ekran, işlemci ve kablosuz haberleşme birimleridir.

### A. İşlemci Birimi

Giyilebilir cihaz tarafında sensörlerden verilerin toplanması, OLED ekranında gösterilmesi ve verilerin mobil cihaza aktarım işlemlerinin yürütülebilmesi için Teensy 3.2 kiti kullanılmıştır. Kart ARM mimarisine sahip, düşük güç tüketimli ve çok sayıda giriş/çıkış yönetebilme yeteneklerine sahip 32 Bit, 72 Mhz, Cortex-M4 tipi işlemci barındırmaktadır. Kartın avantajından biri de küçük boyutlara sahip olmasıdır.

### B. Kablosuz Haberleşme Birimi

Kablosuz haberleşmede ilk aşamada, ZigBEE protokolünün kullanan Xbee cihazları kullanılmıştır. Bu protokol yeni nesil RF (Radyo Frekansı) haberleşmesi olarak adlandırılmaktadır. Düşük güç tüketiminin yanı sıra üç farklı haberleşme ağı kurulabilme özellikleri vardır. Projede başlangıçta Xbee Pro, 2.4 GHz, 60 mW versiyonu kullanılmıştır.

Projenin ikinci aşamasında, Bluetooth haberleşme protokolüne geçilmiş ve BLE 4.0 özelliğindeki, NRF51 modülü kullanılmıştır. Düşük güç tüketimi ile kısa mesafelerde sağlıklı iletim sağlanmıştır. Giyilebilir cihazda toplanan veriler mobil telefona NRF51 üzerinden aktarılmaktadır.

### C. Kablolu Haberleşme

Sistemde I2C haberleşmesi, Teensy 3.2, Max30100, MS4 sensör ve OLED ekran bileşenlerini kontrol etmekte kullanılmaktadır. Bu haberleşme devre üzerinde kısa mesafeli kablo yollarında sıkça tercih edilmektedir. 100 Kbps temel iletim hızını desteklemekte ve iletimin tek bir kablo hattından sağlanmaktadır. Sistemde kullanılan Clock (saat) sinyali sayesinde eşzamanlı bir haberleşme sağlamaktadır.

Bluetooth modül ise seri port (UART Universal Asynchronous Receiver Transmitter) protokolü ile sürülmüştür. UART iletişimde bir kablo hattından veri iletimi (Transmitter-TX) diğer kablo hattından da veri eldesi (Receiver-RX) sağlanmaktadır.

### D. Sensörler-MAX30100 Nabız Ölçer

Bakım personelinin nabız ve oksijen saturasyonu (SpO2) verilerinin elde edilmesinde MAX30100 sensoru kullanılmaktadır. Max30100 sensörü, analogtan sayısala dönüştürücü (DAC), sayısal filtre, işlemci ve çevre elemanlarını içermektedir. Diğer modüllerde olduğu gibi küçük boyutları ve düşük güç tüketimiyle birlikte, kısa mesafelerde tercih edilen I2C kablolu haberleşme protokolünü desteklemektedir. Aynı veri hattı üzerinden hem nabız hem de SpO2 bilgisine ulaşılmıştır. Bu modülün kullanımında ham veriler elde edilerek yazılımsal olarak hata giderimi ve çeşitli düzenlemelerle anlamlı nabız bilgisi elde edilmiştir.

### E. Deri Direnci (GSR Galvanic Skin Response) Sensörü

Giyilebilir cihazı kullanacak olan personelin heyecan, stres ve sinir durumlarında yaşayacağı konsantrasyon kaybı ve buna bağlı bir iş kazası oluşması riski dikkate alınarak deri direnci ölçümü yapılmıştır. İnsan vücudunu kaplayan deri üzerindeki gözeneklerde, ter bezleri sebebiyle sürekli olarak ancak farklı miktarlarda su bulunmaktadır. Ter, bu gözenekler yardımıyla cildi nemlendirir ve hararetin atılmasına katkı sağlar. Ter, iletkenliği arttırarak cilde verilen küçük voltajlardaki elektrik aktivitesinin daha kolay yayılımını sağlar. Deri iletkenliği teknolojisi bu elektriksel yayılımının miktarını algılayarak, duygusal uyarıcının kişide yarattığı şiddeti ölçmeyi amaçlar [15]. Bu hassas ölçüm yöntemlerinden biri Galvanik deri cevabıdır (GSR Galvanic Skin Response).

### F. MS4 Multi-Sensör Modülü

Ortama ait ısı, ışık, nem, CO2, uçucu organik bileşikler (VOC Volatile Organic Compound), ses ve hareket verilerinin toplanmasında kullanılmıştır. Sensörün küçük boyutları ve I2C protokol desteği en büyük avantajlarıdır. Ortama ait parametreler anlık olarak toplanarak, herhangi bir değerde meydana gelecek anormal değişiklikler izlenecektir. Bu sayede çalışma ortamının uygun şartlarda olup olmadığı takip edilerek kullanıcıya bildirilecektir. Ortam parametrelerinden her bir değere ait ham veriler elde edilmiş ve yazılımsal düzenlemeler ve filtreleme sayesinde gürültüden arındırılan veriden anlamlı bilgiler elde edilmiştir.

#### G. Operatör Paneli Birimi-OLED Ekran

Cihazı kullanan personelin gerek kendisine ait sağlık parametrelerini gerekse ortama ait bilgileri takip edebilmesi için 0.96 inch büyüklüğünde, düşük güç tüketimli, I2C haberleşme desteği olan bir renkli bilgilendirme ekranı kullanılmıştır.

#### H. Çevre Elemanları

Uyaran ve bilgilendiricilere ek olarak sistemde kırmızı bir LED ve cihazın açılıp kapatılması için butonlar kullanılmıştır. Sağlık parametreleri anormal bir seyir sergilediğinde LED aktif hale getirilmektedir.

#### I. Şarj Edilebilir Güç Ünitesi

Şarj edilebilir güç ünitesi, giyilebilir cihazı enerjilendirerek çalışmasını sağlamaktadır. Sistemde kullanılan batarya, Lityum Polimer (Li-Po) pil olarak seçilmiştir. Giyilebilir güvenlik sisteminin kontrol ünitesi, yani makinelere entegre edilen ünite ise doğrudan bir elektrik şebekesine bağlıdır, harici batarya bloğu içermez.

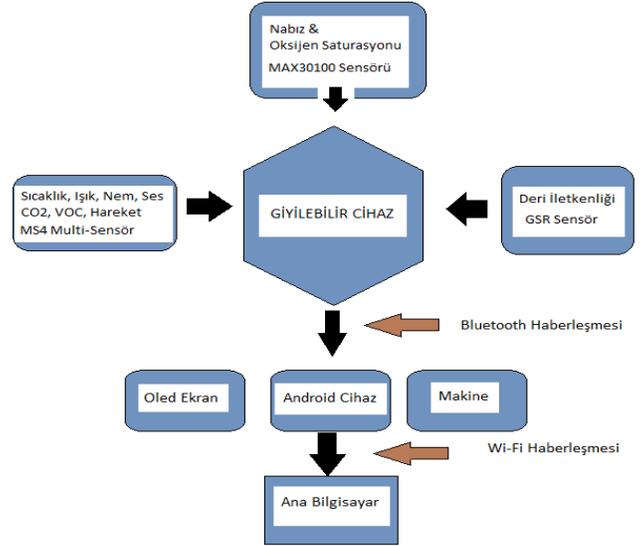

ŞEKİL 4- Genel Akış

### 3. Uygulamanın Gerçekleştirilmesi

Operatörün kullandığı cihaz üzerinden sensörler vasıtasıyla alınan sağlık ve ortam verileri öncelikle OLED ekrana aktarılmakta, devamında kablosuz haberleşme kanalları üzerinden Android mobil cihaza gönderilmektedir. Mobil cihazda bulunan dâhili jiroskopun sensör verileri de toplanarak oluşturulan veri, kablosuz ağ üzerinden bir bilgisayara aktarılmaktadır. Sağlık veya ortam koşullarında anormal bir durum oluşması halinde hem ana bilgisayardan hem de giyilebilir cihaz üzerinde bulunan buton ile makine manuel olarak uzaktan durdurulacaktır. Sistemin bileşenleri ve işlem akışı Şekil 3'de ayrıntılı bir şekilde gösterilmektedir.

Uygulama, Teensy 3.2 işlemci kartı, NRF51 Bluetooth kartı, MAX30100, GSR ve MS4 sensör kartları ve OLED ekranın birleştirilmesiyle gerçekleştirilmiştir. Test uygulamalarında anlamlı sensör değerlerinin toplanması, bu verilerin OLED ekrana ve Bluetooth modül ile bir Android mobil cihaza ve server bilgisayara aktarılması işlemleri ayrı ayrı ve birlikte gerçekleştirilmiştir. Uygulamada I2C haberleşme hattında anlamlı verilerin iletimi ve sağlıklı bir bilgi transferi için donanımsal eklentiler yapılmıştır. I2C haberleşme yolunun hem veri (SDA-Serial Data Line) hem de saat (SCL- Serial Clock Line) hattına Pull-Up yapıda 4.7 KOhm değerinde dirençler bağlanmıştır. Pull-Up direnç yapısının amacı, fabrika ortamlarında yüksek akım tüketen yüklerin neden olduğu yayılım ve iletim gürültülerini bastırarak, giriş voltajının bu her iki hatta da net olarak lojik-1 veya lojik-0 olarak algılanmasının sağlanmasıdır.

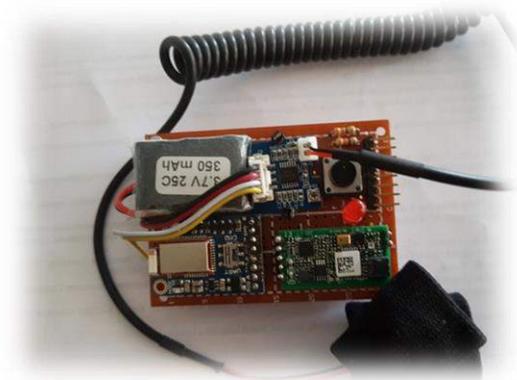

ŞEKİL 5-Sistemin Ön Görünüşü

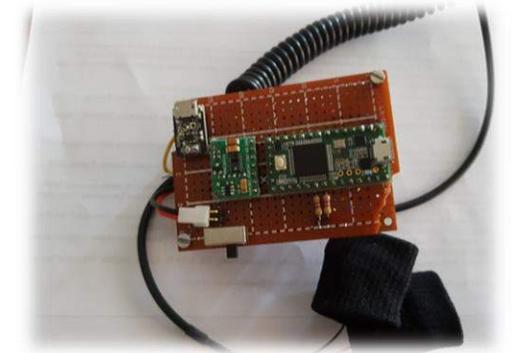

ŞEKİL 6-Sistemin Arka Görünüşü

## 4. Sonuçlar ve Tartışma

Bu uygulamada I2C haberleşmesi ile 100 Kbps veri iletim hızında kullanıcının üzerinde taşıdığı sensörlerden nabız, oksijen saturasyonu, vücut sıcaklığı ve deri iletkenlik değerleri ile ortamın sıcaklığı, ışık, nem ve karbondioksit ($CO_2$) verileri anlamlı şekilde elde edilmiştir. Toplanan veri OLED ekrana, Android mobil telefona ve bilgisayar üzerinde çalışan Windows uygulamasına aktarılarak takip edilmiştir.

Bundan sonraki çalışmada 10 personel üzerinde yapılacak çalışmada sahadan gerçek veriler alınacak, ramak kalalar ve geçmiş iş kazaları incelenerek elde edilen tecrübeler neticesinde personel bilgileri yorumlanacaktır. Şekil 2'de verilen faklı risk kategorilerine ait makine bölgeleri üzerinde bakım yapan personelin sağlık parametreleri ve ortam koşullarına ait bilgiler toplanarak makine öğrenmesi esaslı bir otomatik karar destek sistemi yazılımı geliştirilecektir. Bu karar destek sistemi ile bakım yapan personelin risk haritası çıkartılarak kişinin ciddi bir risk altında olup olmadığı sonucuna varılabilecek ve gerektiği durumlarda çalışma yapılan makine otomatik olarak acil pozisyona geçirilebilecektir.

## 5. Kaynaklar